# Strong Electron-Phonon Coupling Superconductivity Induced by a Low-lying Phonon in IrGe

*Daigorou Hirai\*, Mazhar. N. Ali, and R. J. Cava*

*Department of Chemistry, Princeton University, Princeton, New Jersey 08544, USA*

The physical properties of the previously reported superconductor IrGe and the $Rh_{1-x}Ir_xGe$ solid solution are investigated. IrGe has an exceptionally high superconducting transition temperature ($T_c$ = 4.7 K) among the isostructural 1:1 late-metal germanides $M$Ge ($M$ = Rh, Pd, Ir and Pt). Specific-heat measurements reveal that IrGe has an anomalously low Debye temperature, originating from a low-lying phonon, compared to the other $M$Ge phases. A large jump at $T_c$ in the specific-heat data clearly indicates that IrGe is a strong coupling superconductor. In the $Rh_{1-x}Ir_xGe$ solid solution, a relationship between an anomalous change in lattice constants and the Debye temperature is observed. We conclude that the unusually high $T_c$ for IrGe is likely due to strong electron-phonon coupling derived from the presence of a low-lying phonon.

KEYWORDS: IrGe, Superconductivity, Strong electron-phonon coupling, low-lying phonon, Transition-metal germanide

## Introduction

Strong electron-phonon coupling has been discussed as the origin of the enhancement of the superconducting transition temperature ($T_c$) in a wide variety of materials. In these superconductors, structural instabilities or low-lying phonons are known to enhance the electron-phonon coupling and hence $T_c$. In $KOs_2O_6$ ($T_c$ = 9.6 K), for example, a characteristic low-lying phonon, the so-called rattling mode, originates from the large size-mismatch between the cage-like framework and the inclusion ion, and has been cited as the source of the unusually high transition temperature.[1]

More than 400 compounds of the transition metals ($M$) with metalloids ($X$) (e.g. Si, Ge, P, As, Se, Te etc.) crystallize in the NiAs structure type[2], which in its ideal form has hexagonal symmetry and is made from layers of edge-sharing $MX_6$ octahedra that then share faces from one layer to the next. The orthorhombic MnP-type structure is one of the most common derivatives of the NiAs-type structure. In this case, $M$-$M$ and $X$-$X$ bonds cause contraction of the ideal distances, distortion of the $MX_6$ octahedra, and the formation of intersecting metal chains through the structure.[3] Several 4$d$ and 5$d$-based transition-metal germanides $M$Ge ($M$ = Rh, Pd, Ir and Pt) crystallize in this structure type.[4-6] Superconducting $T_c$'s of 0.96, 4.7 and 0.40 K have been reported for RhGe, IrGe and PtGe, respectively, and PdGe is not known to superconduct above 0.4 K.[7] The $T_c$ of 5$d$-based IrGe is exceptionally high among these isostructural compounds: five times that of RhGe, its 4$d$ analog, and ten times that of the compound based on Pt, Ir's neighbor in the periodic table. This anomalous behavior suggests that a key factor might be present in IrGe that enhances its $T_c$.

Here we report a detailed characterization of the physical properties of IrGe and a comparison to the isostructural $M$Ge compounds with $M$ = Rh, Pd and Pt. Our specific-heat data clearly indicate that IrGe is an $s$-wave, strong-coupling superconductor. The ratio between the superconducting gap and transition temperature evaluated from specific heat data, $2\Delta_0/k_BT_c$ = 5.17, a good measure of strong electron-phonon coupling, is anomalously large. Further, the existence of a low-lying phonon is seen in the normal state specific-heat data, reflected in an anomalously low Debye temperature $\Theta_D$ = 153 K in IrGe compared to its neighboring phases. In the $Rh_{1-x}Ir_xGe$ solid solution, we observe an anisotropic non-linear dependence of lattice constants on composition, suggesting the proximity of a structural instability for IrGe. Although the crystal structures of IrGe and RhGe are very similar, they differ in the extent of the $M$-$M$ bonding in the metal atom framework, which may play a role in the presence of the low-lying phonon. We argue that the proximity of IrGe to a structural instability is reflected in its anomalously high $T_c$, the strong electron-phonon coupling, and its low Debye temperature.

## Experimental

Polycrystalline $M$Ge ($M$ = Rh, Ir, Pd and Pt) samples were prepared by conventional solid-state reaction. A mixture of transition metals and elemental germanium was pelletized and sintered in a sealed quartz tube at 700 °C for 3 days under an atmosphere of Ar gas. The sintered pellet was reground, repelletized, and sintered again at 700 °C for 1 week. For the solid-solution samples $Rh_{1-x}Ir_xGe$ ($0 \leq x \leq 1$), stoichiometric amounts of pre-reacted RhGe and IrGe were mixed and heated at 750 °C for 3 days in a sealed quartz tube under a partial atmosphere of Ar gas. Single crystals of IrGe and RhGe, used for the single crystal X-ray diffraction (XRD) studies, were grown from a Bi flux. Polycrystalline IrGe and RhGe were mixed with Bi in a ratio of $M$Ge: Bi = 1:20 and placed in an alumina crucible in a sealed quartz tube under an atmosphere of Ar gas. The sealed tube was heated to 1050 °C for 5 h and slowly cooled at the rate of 3 °C/h to 600 °C. At 600 °C, the Bi flux was decanted from the $M$Ge crystals by centrifugation. The residual Bi on the surface of crystals was washed away by dilute nitric acid.

---

\* E-mail: dhirai@issp.u-tokyo.ac.jp

The quality of the polycrystalline samples was checked by laboratory XRD using Cu $K\alpha$ radiation on a Bruker D8 Focus diffractometer with a graphite monochromator. Single crystal XRD was conducted for IrGe and RhGe on a Bruker APEX II diffractometer using Mo $K\alpha$ radiation. Synchrotron powder XRD data for IrGe were collected at beamline 11-BM at the Advanced Photon Source. The crystal structures of IrGe and RhGe were refined from the single crystal data using SHELXL-97 implemented through WinGX. The crystallographic cell parameters of Rh$_{1-x}$Ir$_x$Ge ($0 \leq x \leq 1$) powders were refined using the FULLPROF program.[8] Electrical resistivity and specific-heat measurements were performed by using a PPMS (Physical Properties Measurement System, Quantum Design) instrument.

**Results**

The superconducting transition in IrGe was examined by resistivity and specific-heat measurements, as shown in Fig. 1. The resistivity reaches the zero resistance state, hallmarking superconductivity, at 5.3 K. The superconducting transition is systematically suppressed under magnetic field. The upper critical field $\mu_0H_{c2}(T)$ of IrGe was evaluated from the mid-point of the resistive transitions, and is summarized in the inset of Fig 1(a). The linearly extrapolated $\mu_0H_{c2}(T)$ to $T = 0$ gives an estimation of $\mu_0H_{c2}(0) = 0.82$ T, which in turn yields a Ginzburg-Landau coherence length of approximately $\xi_{GL} = 200$ Å.

Evidence for the bulk nature of the superconductivity in IrGe was obtained from the large specific-heat jump at $T_c$, shown in Fig. 1(b). The electronic specific-heat under zero applied field $C_{el}(T)$ can be estimated by subtracting the lattice contribution of specific-heat obtained from the normal state specific-heat $C_N(T)$. The normal state specific-heat is estimated by suppressing the superconducting state under a magnetic field of $\mu_0H = 2$ T, as shown in Fig. 2(a). The fitting with $C_N(T) = \gamma T + \beta T^3$ gives an estimation of $\gamma = 3.53$ mJ/mol K$^2$ and $\beta = 0.997$ mJ/mol K$^4$. The $\gamma$ value obtained is moderately low, as expected for 5$d$ intermetallic compounds. Taking the entropy balance around the transition into account, a $T_c = 5.17$ K is obtained from the specific-heat data, which is slightly higher than the reported value of $T_c = 4.7$ K,[7] and consistent with our resistivity data. The very sharp specific heat transition indicates that the sample is homogeneous and of high quality.

The temperature dependence of $C_{el}$ below $T_c$ clearly shows that IrGe is a strong-coupling superconductor with an isotropic energy gap. The specific-heat jump at $T_c$, $\Delta C/\gamma T_c = 3.04$, is much larger than the value of 1.42 expected for the BCS weak coupling limit. In addition, $C_{el}(T)$ decreases exponentially below $T_c$, implying an isotropic superconducting gap. Quantitative analysis for $C_{el}(T)$ was performed by the fitting with the semiempirical approximation, the so-called $\alpha$ model, $C_{el}(T) = A \exp(-\Delta_0/k_BT)$, where $k_B$ and $\Delta_0$ are the Boltzmann constant and the superconducting gap at 0 K, respectively.[9] This equation allows for varying the coupling strength $\alpha = \Delta_0/k_BT_c$, instead of fixing it at the BCS weak-coupling limit, $\alpha = 1.76$. The obtained coupling strength $\alpha = 2.57$ (solid line) gives an excellent fit, while the BCS weak-coupling limit (dotted line) fails to describe the data. The obtained $2\Delta_0/k_BT_c = 5.17$, which is much larger than the weak coupling BCS value of 3.52, clearly indicates that IrGe is a strong coupling superconductor. Only a limited number of strong coupling superconductors are known to have an $2\Delta_0/k_BT_c$ value as high as 5; the pyrochlore osmates[1], Chevrel phases[10], Pb-Bi alloys[11] and SrPt$_3$P[12] are examples.

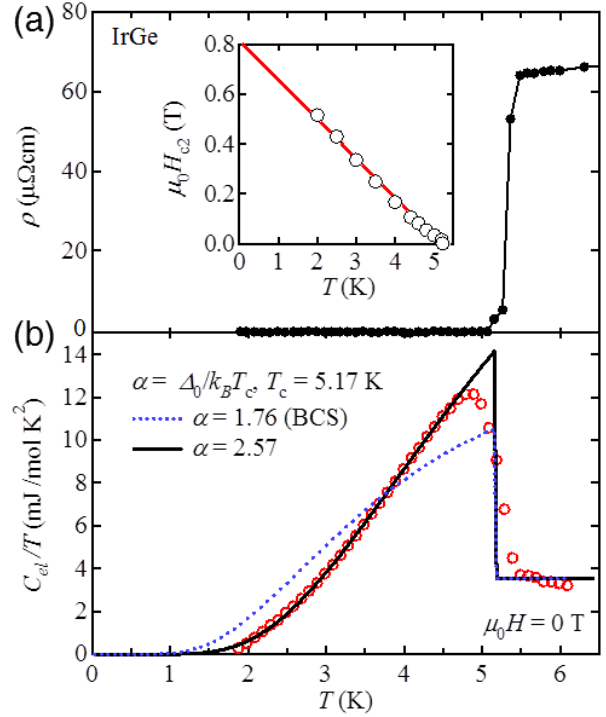

Fig. 1. The temperature dependence of (a) the electrical resistivity ($\rho$) and (b) the electronic specific heat divided by temperature ($C_{el}/T$) for IrGe around the superconducting transition. The solid line is a fit to $C_{el}/T(T)$ based on an empirical fitting function with variable $\alpha$,[9] and the dotted line shows that of weak-coupling BCS model. Inset in (a) shows the temperature dependence of the upper critical field $\mu_0H_{c2}(T)$ obtained from resistivity data under magnetic field.

The unusual properties of IrGe were also observed in its normal state specific-heat data, as shown in Fig. 2(a). As described above, the normal state specific-heat at low temperatures can be fitted as $C_N(T) = \gamma T + \beta T^3$. The fitting of our data below 6 K for RhGe, IrGe, PdGe, and PtGe yields electronic specific-heat coefficients ($\gamma$) of 3.55, 3.53, 1.45, and 1.39 mJ/mol K$^2$, respectively. The values of $\gamma$ for the isoelectronic compounds in this structural type, i.e RhGe and IrGe, and PdGe and PtGe, are almost equivalent irrespective of whether the material is based on the 4$d$ or 5$d$ metal; $\gamma$ reflects the density of states (DOS) at the Fermi energy ($E_F$) renormalized by the electron-phonon coupling (see below). The Debye temperatures of 303, 299, 158, and 274 K are also obtained from the fitting for RhGe, PdGe IrGe, and PtGe, respectively ($\Theta_D = (12\pi^4NR/5\beta)^{1/3}$). The Debye temperature obtained for IrGe is anomalously low compared with the other compounds in the family. In a simple harmonic oscillator, an atom with larger mass $M$ is expected to give a lower characteristic frequency $\omega$, as $\omega$ is proportional to $M^{-1/2}$. In a solid, this will be reflected in the Debye temperature. The Debye temperatures for the 4$d$-based compounds with comparable masses in this family, RhGe and PdGe, are almost the same. In contrast, however, the Debye temperature for IrGe is approximately half of that of PtGe, a compound with approximately the same molar mass. Comparison of the specific-heats up to 200 K [Fig. 2(b)] reveals that the origin of the anomalously low Debye temperature of IrGe is a low-lying phonon. At high temperature, the specific-heats of both IrGe and PtGe saturate at the ideal value of 6$R$ per mol of formula unit, as expected by the Dulong-Petit law. The specific-heat of IrGe saturates at slightly higher temperature than that of PtGe, however, in apparent contradiction with the much lower Debye temperature in IrGe. In order to clarify the origin of the low Debye temperature of IrGe, the difference of $C/T(T)$ between IrGe and PtGe is plotted in the inset of Fig. 2(b). There is a large extra contribution to the specific-heat at low temperatures ($T <$ 25 K) in IrGe; this yields the very low Debye temperature evaluated from the low temperature fitting.

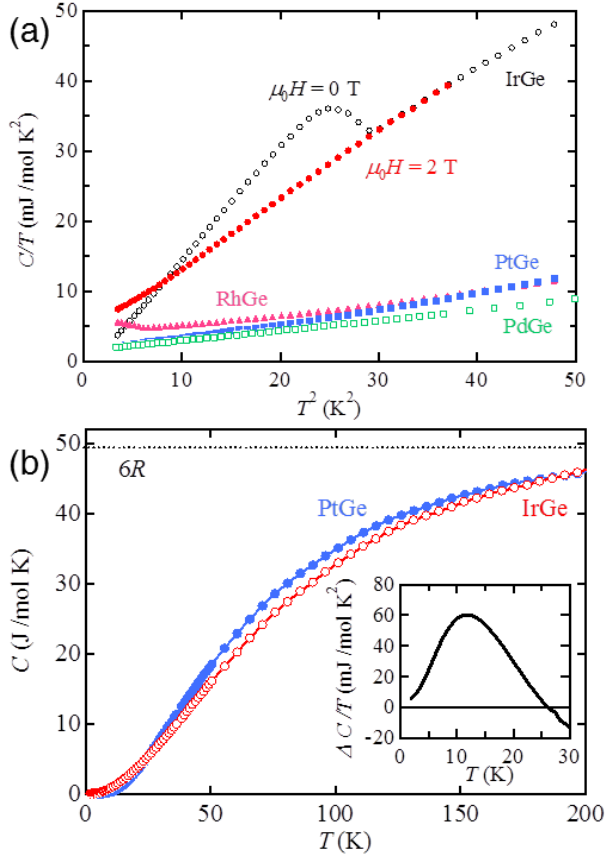

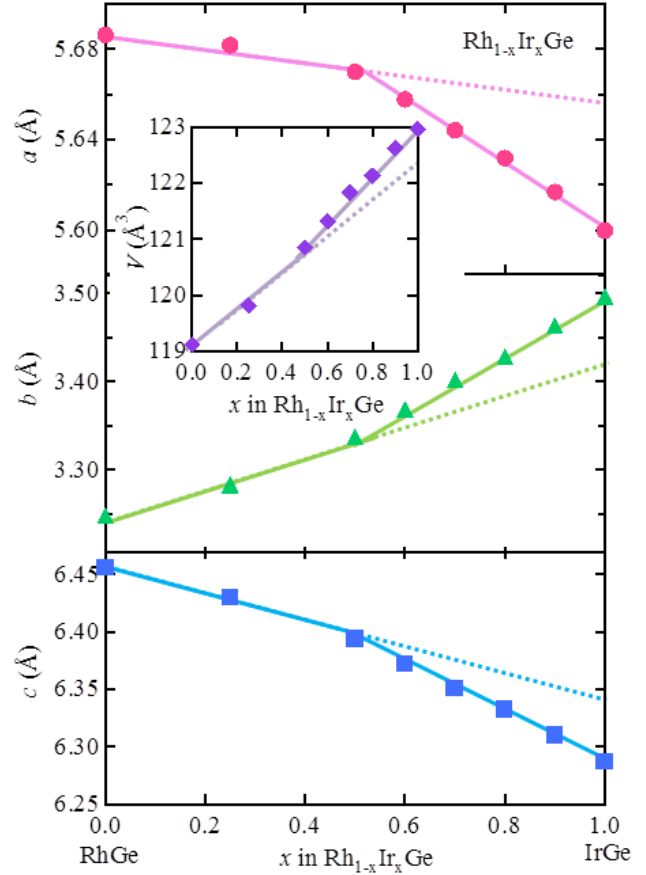

Fig. 2. (a) Specific heat/Temperature ($C/T$) as a function of temperature squared ($T^2$) for $M$Ge ($M$ = Ir, Rh, Pt, and Pd) at low temperature. (IrGe $\mu_0 H$ = 2 T, filled circles, IrGe $\mu_0 H$ = 0 T open circles; RhGe, filled triangles, PtGe, filled squares; and PdGe open square. For RhGe, IrGe and PtGe, measurements are in zero applied field) (b) Comparison of specific heats $C(T)$ of IrGe (in $\mu_0 H$ = 2 T) and PtGe (in $\mu_0 H$ = 0 T) below 200 K. Inset shows $C/T$ of IrGe subtracted from that of PtGe showing the rapid increase of $C/T$ in IrGe below 30 K.

Fig. 3. The lattice constants (main panels) and cell volume (insert) of $Rh_{1-x}Ir_xGe$. The lines are a guide to the eye. The error bars are smaller than the plotted points.

Although RhGe is isostructural and isoelectronic to IrGe, there are significant differences in their $T_c$s and superconducting properties. In order to examine the relationship between very strong coupling superconductivity and the low-lying phonon in IrGe, the solid solution of IrGe with RhGe was investigated. The lattice parameters for the $Rh_{1-x}Ir_xGe$ ($0 \leq x \leq 1$) solid solution are shown in Fig. 3. Although IrGe and RhGe have the same structure type (see below), pronounced features can be seen in the dependence of lattice parameters against the Ir content $x$. A systematic expansion of the unit cell volume is observed, as expected from the larger atomic radius of Ir compared with Rh (inset of Fig. 3). Instead of an isotropic expansion of the unit cell dimensions, however, one cell parameter grows while the other two shrink dramatically. Further, changes in the rate that the cell parameters vary with composition begin at $x$ = 0.5 in $Rh_{1-x}Ir_xGe$ and extend to higher Ir contents.

The kink in the $x$ dependence of lattice parameters near $x$ = 0.5 was also observed in the specific-heat data, as shown in Fig. 4(b) and 4(c). The Debye temperatures and electronic specific coefficients were obtained from fittings to the low temperature (2 K ≤ $x$ ≤ 6 K) specific-heat data for $Rh_{1-x}Ir_xGe$. The Debye temperature decreases monotonically on increasing Ir content and is smallest at the IrGe limit. Given that the origin of the very low Debye temperature in IrGe is the existence of low-lying phonon, the reduction of Debye temperature from RhGe to IrGe indicates the gradual evolution of the low-lying phonon toward the IrGe limit. The similar $x$ dependence of the lattice parameters and Debye temperature strongly suggests a close relationship between the low-lying phonon mode and the crystal structure. In contrast, $\gamma$ varies very little in the series between RhGe and IrGe [Fig. 4(b)]. The virtually identical $\gamma$ values (i.e. 3.55 mJ/mol-$K^2$ in RhGe and 3.53 mJ/mol-$K^2$ in IrGe) exclude the possibility that the higher $T_c$ for IrGe can be due to a larger DOS at $E_F$. (Due to the higher electron-phonon coupling for IrGe, its DOS at $E_F$ must actually be smaller than that of RhGe to give nearly identical $\gamma$ values.) Therefore the unusual enhancement of $T_c$ in IrGe likely originates from the reinforcement of the pairing interaction through enhanced electron-phonon coupling, rather than due to the enhancement of the DOS at $E_F$. The broad superconducting transition of RhGe (inset of Fig. 4) makes the precise determination of $T_c$ and the specific-heat jump difficult, but $\Delta C/\gamma T_c$ is clearly much smaller than that of IrGe.

*E-mail: dhirai@issp.u-tokyo.ac.jp

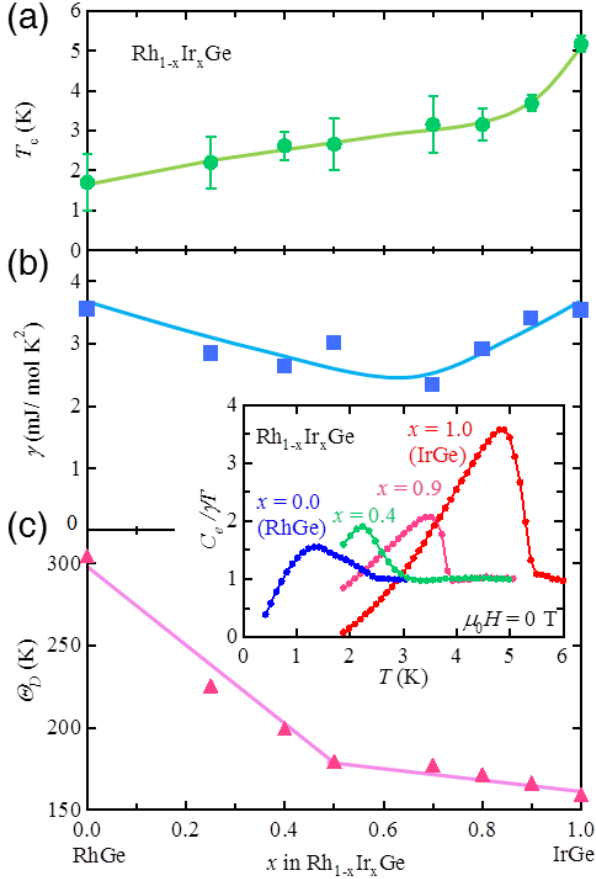

Fig. 4. Variation of the measured $T_c$ (a), electronic contribution to the specific heat (b), and Debye temperature (c) for materials in the $Rh_{1-x}Ir_xGe$ solid solution. The inset shows the temperature dependent specific heat through the superconducting transitions for several $Rh_{1-x}Ir_xGe$ samples. The lines are guides to the eye.

The variation of $T_c$ in the $Rh_{1-x}Ir_xGe$ solid solution is shown in Fig. 4(a). The $T_c$ of $Rh_{1-x}Ir_xGe$ increases smoothly up to $x = 0.9$ and has a larger incremental increase on going to IrGe. Even though the low-lying phonon evolves gradually from RhGe to IrGe, the very large specific-heat jump at $T_c$ ($\Delta C/\gamma T_c$), a hallmark of strong electron-phonon coupling, is observed only at pure IrGe. Thus it appears that the electron-phonon coupling becomes very strong only when the low-lying phonon becomes more dominant at IrGe. We note that the almost identical $\gamma$ values for RhGe and IrGe appear to be in contradiction with the fact that strong electron-phonon coupling is expected to enhance $\gamma$. Considering that 5$d$ compounds usually have smaller DOS at $E_F$ than 4$d$ compounds due to their largely extended $d$-orbitals, the concurrent reduced DOS of IrGe and the enhancement of $\gamma$ due to strong electron-phonon coupling may result in the comparable $\gamma$'s for RhGe and IrGe.

The detailed inspection of the single crystal and high resolution synchrotron powder diffraction data for RhGe and IrGe confirmed that both have the orthorhombic MnP-type structure with no long-range structural distortions, consistent with previous lower resolution studies[4,5]; the structures refined from the single crystal data are presented in Table I.[13] The cell dimensions differ anisotropically on going from RhGe to IrGe, so the cell dimension changes are not simply a reflection of hard-sphere-like atomic sizes. Comparison of the interatomic distances in the structures shows that the major differences involve the three-dimensional connectivity of the metal chains. It is not clear without detailed electro-structural modeling how the subtle differences in the structures impact the strength of the electron phonon coupling. However it may be that the low-lying phonon in IrGe arises from an incipient instability of the Ir-metal framework, which is anomalously smaller than expected from atomic radii, and may therefore have a high susceptibility to deformations leading to lower symmetry.

Table I. Refined structural parameters of RhGe and IrGe from single crystal XRD.

| RhGe | Pnma (No. 62) $wR_2 = 0.0211, R_1 = 0.0101$ | | | |
|---|---|---|---|---|
| $a = 5.704(2)$ Å, $b = 3.215(1)$ Å $c = 6.504(2)$ Å | | | | |
| Atom | site | $x$ | $Y$ | $z$ |
| Rh | 4c | 0.00339(4) | 1/2 | 0.20193(4) |
| Ge | 4c | 0.19144(6) | 1/2 | 0.93977(5) |
|  |  | $U_{11}$ | $U_{22}$ | $U_{33}$ |
| Rh |  | 0.0023(2) | 0.0051(2) | 0.0035(2) |
| Ge |  | 0.0030(2) | 0.0044(2) | 0.0041(2) |
| IrGe | Pnma (No. 62) $wR_2 = 0.0267, R_1 = 0.0150$ | | | |
| $a = 5.605(1)$ Å, $b = 3.484(1)$ Å $c = 6.296(1)$ Å | | | | |
| Atom | site | $x$ | $y$ | $z$ |
| Ir | 4c | 0.00349(4) | 1/2 | 0.20305(4) |
| Ge | 4c | 0.1875(1) | 1/2 | 0.9263(1) |
|  |  | $U_{11}$ | $U_{22}$ | $U_{33}$ |
| Ir |  | 0.0011(2) | 0.0044(2) | 0.0023(2) |
| Ge |  | 0.0019(4) | 0.0030(3) | 0.0032(3) |

In conclusion, we have investigated the physical properties of the superconductor IrGe and the related compounds $M$Ge ($M$ = Rh, Pd and Pt). The very large specific-heat jump at $T_c$ and the exponential decay of $C_{el}/T$ below $T_c$ clearly indicate that IrGe is a strong-coupling superconductor with an isotropic gap. Compared with the isostructural $M$Ge compounds, an anomalously low Debye temperature is observed in IrGe, originating from a low-lying phonon. The superconducting properties of the $Rh_{1-x}Ir_xGe$ solid solution show that the electron-phonon coupling is strongest in IrGe. We attribute the origin of the very strong electron-phonon coupling to an incipient structural instability in that phase. The unusually high $T_c$ superconductivity in IrGe and its associated soft phonon and strong electron-phonon coupling appear to be worthy of further experimental and theoretical consideration.

## Acknowledgments

This work was supported by the AFOSR MURI on superconductivity, grant FA9550-09-1-0953. Use of the APS at Argonne National Laboratory was supported by the U. S. DOE, Office of Science, Office of Basic Energy Sciences, under Contract No. DE-AC02-06CH11357.

* E-mail: dhirai@issp.u-tokyo.ac.jp